\documentclass[proceedings, preprint]{rmaa}

\usepackage{paralist}

\usepackage{psfrag,color}
\usepackage{amsmath}

\SetYear{2009}
\SetConfTitle{Magnetic Fields in the Universe II} 

\title{Statistics of Centroids of Velocity} 

\author{A. Esquivel\altaffilmark{1} and A. Lazarian\altaffilmark{1} }

\altaffiltext{1}{Instituto de Ciencias Nucleares, Universidad Nacional
  Aut\'onoma de M\'exico, Apartado Postal 70-533, M\'exico D. F. 04510,
  M\'exico (esquivel@nucleares.unam.mx).}
\altaffiltext{2}{Department of Astronomy, University of
  Wisconsin-Madison, 475 N. Charter St., Madison, WI 53706, USA
  (lazarian@astro.wisc.edu)} 
\shortauthor{Esquivel \& Lazarian}
\shorttitle{Velocity Centroids}

\listofauthors{A. Esquivel \& A. Lazarian}
\indexauthor{Esquivel, A.}
\indexauthor{Lazarian, A.}
\abstract{We review the use of velocity centroids statistics to
  recover information of interstellar turbulence from 
  observations. Velocity centroids have been used for a long time now
to retrieve information about the scaling properties of the turbulent
velocity field in the interstellar medium. We show that, while
they are useful to study subsonic turbulence, they do not trace the
statistics of velocity in supersonic turbulence, because they are
highly influenced by fluctuations of density. We show also that for
sub-Alfv\'enic turbulence (both supersonic and subsonic) two-point
statistics (e.g. correlation functions or power-spectra) are
anisotropic. This anisotropy can be used to determine the direction of
the mean magnetic field projected in the plane of the sky.
}  

\resumen{
Hacemos un resumen del uso de estad\'\i sticas de mapas de centroides de
velocidad para obtener informaci\'on de la turbulencia
del medio interestelar a partir de observaciones. Los centroides de velocidad
han sido usados por mucho tiempo para obtener informaci\'on de las
propiedades de escalamiento de la velocidad turbulenta en el medio
interestelar. Mostramos que, mientras que los mapas de centroides de
velocidad son \'utiles para el estudio de turbulencia subs\'onica, no
lo son para turbulencia supers\'onica, porque son influenciados por
fluctuaciones de densidad. Mostramos tambi\'en que para turbulencia
sub-Alfv\'enica (tanto subs\'onica como supers\'onica) las 
estad\'\i sticas de dos puntos (como las funciones de correlaci\'on o los
espectro de potencias) muestran anisotrop\'\i a. Esta anisotrop\'\i a puede
ser aprovechada para determinar la direcci\'on del campo magn\'etico
proyectado en el plano del cielo.
}

\addkeyword{MHD}
\addkeyword{turbulence}
\addkeyword{ISM: general}
\addkeyword{ISM: structure}
\addkeyword{radio lines: ISM}

\begin{document}
\maketitle

\section{Introduction}
\label{sec:intro}

It is well known that the interstellar medium (ISM) is turbulent. Such
turbulence is magnetized and expands over several scales, ranging from au to kpc
(\citealt{Lar92}; Armstrong, Rickett \& Spangler 1995; Deshpande,
Dwarakanath \& Goss 2000;
\citealt{2001ApJ...551L..53S}, Lazio et al. 2004). Understanding of this magnetic
turbulence is of great importance for key astrophysical processes, from star formation to diffusion of heat and cosmic rays, we refer the reader to
recent reviews on the subject \citet{2004ARA&A..42..211E}, McKee \& Ostriker (2007).

Observations of line-widths
\citep{Lar81,Lar92,1984ApJ...277..556S,scalo87} and of the centroids
of spectral lines \citep{VH51,M58,KD85,DK85,MB94,OC87} have been used
for well over half a century to study turbulence in the ISM.
An important measure that one can hope to retrieve from spectral line
data is the power-law index of the underlying velocity field. However,
the shape of spectral lines does not depend solely on the velocity
field, but on the density of emitting material as well. The separation
of the two contributions has proven to be a formidable (and in some
aspects still an open) problem \citep[see reviews by][]{L06,L08}. 

In parallel with the increasing number and quality of observations,
there have been substantial theoretical and numerical
advances in our understanding of compressible magnetohydrodynamic
(MHD) turbulence (see Cho \& Lazarian 2003 and also reviews Cho,
Lazarian \& Vishniac (2003) and \citet{CL05}). Therefore comparison
between theory, numerics and observations has become essential.

Since turbulence is essentially an stochastic process, statistical
methods are necessary for its study. Studies of correlations can be obtained in
real space using correlation or structure functions, but also in Fourier space using
spectra. Wavelets, which sometimes are preferable for handling of the real observational data, combine properties of both correlation functions and spectra and
can be related to both of them. 
A well-known example of of wavelets,
the $\Delta$-variance, has been successfully used to retrieve
the same information yielded by two point statistics
(i.e. power-spectra or structure functions)
\citep{SBHOZ98,MLO00,BSO01,OML02}. 
However, neither of the measures above is capable
of separating velocity and density contributions to the spectral
lines. The latter separation should be done on the basis of
theoretical understanding of the statistical properties of the
Doppler-shifted spectral lines\footnote{An example of empirical
  approach to the problem is the use of the Principal Component
  Analysis (PCA) discussed, for instance, in Heyer \& Brayn (2004). We
  feel that this powerful approach does not show all its strength
  being isolated from theory. For instance, it is known that for
  shallow density spectrum the statistics of
  Position-Position-Velocity (PPV) data cubes inseparably depends on
  both spectra of velocity and density (Lazarian \& Pogosyan
  2000). This effect has not been demonstrated so far within the PCA
  approach.}. 

Recent years have been marked by a sharp increase of interest in
statistical techniques of analysis of observations of astrophysical
turbulence. We can mention in this respects two web sites by Alyssa
Goodman, namely ``Taste Tests''\footnote{www.cfa.harvard.edu/$\sim$agoodman/newweb/tastetests.html} where
different comparisons of 
the numerical simulations and observations are presented, and
``Astronomical Medicine'' site\footnote{am.iic.harvard.edu} where
application of sophisticated medical analysis software are applied to
astronomical data. In addition, we may mention new ways of analyzing
column densities, such as ``Genus'' (Lazarian 1999, Lazarian, Pogosyan \&
Esquivel 2002, Kim \& Park 2007, Chepurnov et al. 2008), ``Bispectrum''
and ``Bicoherence'' (Lazarian 1999, Lazarian, Kowal \& Beresnyak 2008,
Burkhart et al. 2009). A discussion of those, is, however, beyond the
scope of our present publication, which deals with a way of extracting
the statistics of velocity from observed Doppler-shifted spectral
lines.

Among new theoretically-motivated way of recovering velocity statists
we can mention ``Velocity Channel Analysis''
\citep[VCA;][]{LP00,LPVP01,LPE02,2003MNRAS.342..325E,LP04}\footnote{The ``Spectral Correlation Function'' (SCF) \citep*{RGWW99,PRG01} is another new measure, which, however,
under close examination differ from the measure in VCA only by its normalization. The advantage of the normalization adopted in the VCA is that the observations can be described in terms of underlying correlations of velocity and density, which is not
the case for the normalization adopted for the SCF.}, ``Modified
Velocity Centroids'' \citep[MVCs;][the first three hereafter LE03 and
  EL05, OELS06,
  respectively]{LE03,EL05,2006A&A...452..223O,2007MNRAS.381.1733E},
and the  ``Velocity Coordinate Spectrum'' \citep[VCS;][]{LP06,LP08}. This paper
explains when velocity centroids, including MVCs are capable of measuring the statistical properties of the underlying velocity turbulence.

The layout of this paper is as follows. In \S\ref{sec:basics} we
present the basic statistical toolds used. We will review some of our
work about the retrieval of velocity statistics from velocity centroids
in \S\ref{sec:tracers}. A special emphasis on the anisotropies in
the statistics that result from the presence of a magnetic field, and
a discussion of how can they be used to reveal the direction of he
mean magnetic field is presented in \S\ref{sec:anisotropy}. Finally we
provide with a brief summary in \S\ref{sec:summary}.

\section{Two-point statistics}
\label{sec:basics}

Two-point statistics, such as correlation/structure functions and
power spectra are the simplest, and most widely used method to characterize
turbulence. The (second-order) correlation function of a quantity
$f(\mathbf{x})$ is defined as:
\begin{equation}
SF(\mathbf{r})=\left\langle \left[ f(\mathbf{x})-
f(\mathbf{x}+\mathbf{r})\right]^2 \right\rangle,
\label{eq:SF}
\end{equation}
where $\mathbf{r}$ is the ``lag'', and $\langle...\rangle$ denotes an ensemble
average over all the space ($\mathbf{x}$). The correlation function
(or auto-correlation function)
\begin{equation}
CF(\mathbf{r})=\left\langle  f(\mathbf{x})\cdot
f(\mathbf{x}+\mathbf{r})\right\rangle,
\label{eq:CF}
\end{equation}
can be easily related structure function as
${SF(\mathbf{r})=2\left[CF(\mathbf{0})- CF(\mathbf{r})\right]}$, where
$CF(\mathbf{0})$ is the variance of $f(\mathbf{x})$. The
power spectrum can be defined as the Fourier transform of the
correlation function,
$P(\mathbf{k})=\mathcal{F}\left[CF(\mathbf{r})\right]$, where $k$ is
the wave-number.

The usefulness of these type of functions lies in the fact that
they have a power-law behavior in the so-called inertial range.
Energy is injected at large scales, and cascades down without losses
down to the scales at which (viscous) dissipation occurs. The inertial
range is precisely between these two scales. For instance, the
Kolmogorov model of hydrodynamical (and incompressible) turbulence
predicts that the difference in velocities at different points in the
fluid increases on average as the cubic root of the separation
($\vert\delta_v\vert\propto l^{1/3}$). This famous 
scaling results in a  structure function
$SF(r)\propto r^{2/3}$, and a (three dimensional) power spectrum
$P(k)\propto k^{-11/3}$. Notice that the Kolmogorov model assumes isotropic
turbulence (i.e. we have replaced $\mathbf{r}$ by
$r=\vert \mathbf{r}\vert$,and  $\mathbf{k}$ by
$k=\vert \mathbf{k}\vert$).
But, turbulence is not isptopic in general. In particular, by
introducing a preferential direction of motion, the magnetic field
that threads the ISM makes the turbulent cascade anisotropic.
There are, however, some statistical measures that
are not very sensitive to such anisotropy \citep[for instance VCA,
  see][]{2003MNRAS.342..325E}. And in fact, it is customary to average
correlation and structure functions in shells (or annuli) of equal
separation (or average the power spectrum in wave number) effectively
reducing the statistics to one dimension.
In the following section (\S\ref{sec:tracers}) we will do such
averaging procedure, but we will come back to discuss how can these
anisotropies be exploited to learn something about the magnetic field
in \S\ref{sec:anisotropy}.

\section{Tracing the statistics of velocity with centroids}
\label{sec:tracers}

Several studies, with varying degrees of success, have been made to
obtain the spectral index of turbulence (power-law index of the
velocity spectrum, correlation or structure function).
However, many of them are restricted to ionized media
(interestellar scintillations for instance), and more importantly they
are only sensitive to fluctuations in density.
While, density fluctuations are a natural
result of a turbulent cascade, one has to make the leap from the 
observed fluctuations of density to a dynamical quantity predicted by
theoretical models, such as velocity or magntetic field. It is
therefore very desirable to obtain directly the spectral index of
velocity from observations.

Evidently, the Doppler-shifted spectral lines contain
information about the velocity. However, they are also affected by
density fluctuations, and one has to be careful to
interpretate the statistics drawn from observations, in particular
results from 2D maps of velocity centroids.
Velocity centroids are ussually defined as \citep{VH51,M58}:
\begin{equation}
C(\mathbf{X})=\frac{\int v_z\, I_{\mathrm{line}}\left(\mathbf{X},v_z\right)\, \mathrm{d}v_z}{\int I_{\mathrm{line}}\left(\mathbf{X},v_z\right)\,\mathrm{d}v_z},
\label{eq:cent1}
\end{equation}
where $I_{\mathrm{line}}$ is the line intensity at a position
$\mathbf{X}=(x,y)$ in the plane of the sky, at
line of sight (LOS) velocity $v_z$.
The integration limits are defined by the extent of velocities
covered by the object.
{\it If the medium is optically thin, and the  emissivity is 
proportional to the  density} (i.e. H\small{I} for instance), one can
replace the velocity integrals by integrals over the actual LOS
(chosen here to coincide with the $z-$axis, see LE03, EL05): 
\begin{equation}
C(\mathbf{X})=\frac{\int v_z\left(\mathbf{x}\right)\, \rho\left(\mathbf{x}\right) \, \mathrm{d}z}{\int \rho\left(\mathbf{x}\right)\,
  \mathrm{d}z},
\label{eq:cent}
\end{equation}
where $\mathbf{x}=(x,y,z)$.
One could construct the structure function of these centroids, but the
denominator of eq.~(\ref{eq:cent}) makes the algebra a bit messy, and
does not provide a significant difference over ``unnormalized
centroids'' (see LE03),
\begin{equation}
S(\mathbf{X})=\int v_z\, I_{\mathrm{line}}(\mathbf{X},v_z)\,dv_z=\alpha \int v_z(\mathbf{x})\, \rho(\mathbf{x})\,dz\,.
\label{eq:ucent}
\end{equation} 

Replacing $\mathbf{x_1}=\mathbf{x}$ and
$\mathbf{x_2}=\mathbf{x}+\mathbf{r}$, and similarly
$\mathbf{X_1}=\mathbf{X}$ and  $\mathbf{X_2}=\mathbf{X}+\mathbf{R}$, 
the structure function of centroids can be written as:
\begin{equation}
\left\langle
\left[ S(\mathbf{X_{1}})-S(\mathbf{X_{2}})\right]^{2}\right\rangle
=\alpha^{2}\iint dz_{1}dz_{2}
\left[D(\mathbf{r})-\left.D(\mathbf{r})\right|_{\mathbf{X_{1}}=\mathbf{X_{2}}}\right],
\label{eq:S1S22A}
\end{equation}
where
\begin{equation}
D(\mathbf{r})=\left\langle\left[\rho\left(\mathbf{x_1}\right)v_z\left(\mathbf{x_1}\right)-\rho\left(\mathbf{x_2}\right)v_z\left(\mathbf{x_2}\right) \right]^{2}\right\rangle.
\label{eq:DA}
\end{equation}
The notation $|_{\mathbf{X_{1}}=\mathbf{X_{2}}}$ indicates that the
integral is to be computed for a zero distance between 
$\mathbf{X_{1}}$ and $\mathbf{X_{2}}$ for the second term (i.e. varying
only in $z$).
Writing the density and velocity fields as a mean plus a
fluctuating part ($\rho=\rho_0+\tilde{\rho}$, $v_z=v_0+\tilde{v}_{z}$),
and approximating the fourth order moments as a combination of second
order moments (see LE03, EL05), the structure function of unnormalized
centroids becomes: 
\begin{equation}
\left\langle
\left[ S(\mathbf{X_{1}})-S(\mathbf{X_{2}})\right]^{2}\right\rangle
\approx I1(\mathbf{R})+I2(\mathbf{R})+I3(\mathbf{R})+I4(\mathbf{R}),
\label{eq:S1S2Dec}
\end{equation}
where 
\begin{subequations}
\begin{eqnarray}
I1(\mathbf{R})=&\alpha^{2}\left\langle v_{z}^{2}\right\rangle
\iint dz_{1}dz_{2}
\left[d_{\rho}(\mathbf{r})-\left.d_{\rho}(\mathbf{r})\right|_{\mathbf{X}_{1}=\mathbf{X}_{2}}\right],
\label{eq:I1}\\
I2(\mathbf{R})=&\alpha^{2}\left\langle \rho^{2}\right\rangle
\iint dz_{1}dz_{2}
\left[d_{v_{z}}(\mathbf{r})-\left.d_{v_{z}}(\mathbf{r})\right|_{\mathbf{X}_{1}=\mathbf{X}_{2}}\right],
\label{eq:I2}\\
I3(\mathbf{R})=&-\frac{1}{2} \alpha^{2}\iint
dz_{1}dz_{2}\left[d_{\rho}(\mathbf{r})d_{v_{z}}(\mathbf{r}) \nonumber\right.\\
& \quad - \left.\left.d_{\rho}(\mathbf{r})\right|_{\mathbf{X_{1}}=\mathbf{X_{2}}}\left.d_{v_{z}}(\mathbf{r})\right|_{\mathbf{X_{1}}=\mathbf{X_{2}}}\right],
\label{eq:I3}\\
I4(\mathbf{R})=&\alpha^{2}\iint dz_{1}dz_{2}\left[c(\mathbf{r})-\left.c(\mathbf{r})\right|_{\mathbf{X}_{1}=\mathbf{X}_{2}}\right].
\label{eq:I4}
\end{eqnarray}
\label{eq:I1234}
\end{subequations}
We have made use of the 3D structure functions of the density and
of the LOS velocity: 
\begin{subequations}
\begin{eqnarray}
d_{\rho}(\mathbf{r})&=&
\left\langle\left[\tilde{\rho}\left(\mathbf{x_1}\right)-\tilde{\rho}\left(\mathbf{x_2}\right)\right]^{2}\right\rangle,
\label{eq:drho}\\
d_{v_z}(\mathbf{r}) &=&
\left\langle\left[\tilde{v}_{z}\left(\mathbf{x_1}\right)-\tilde{v}_{z}\left(\mathbf{x_2}\right)
  \right]^{2}\right\rangle, 
\label{eq:dvz}
\end{eqnarray}
\end{subequations}
and the  remaining density-velocity cross-correlations have been
grouped into
\begin{equation}
c(\mathbf{r})=2\left\langle
\tilde{v}_{z}(\mathbf{x_{1}})\tilde{\rho}(\mathbf{x_{2}})\right\rangle
^{2}-4\rho_{0}\left\langle\tilde{\rho}(\mathbf{x_{1}})\tilde{v}_{z}(\mathbf{x_{1}})\tilde{v}_{z}(\mathbf{x_{2}})\right\rangle.
\label{eq:c}
\end{equation}
In the decomposition proposed in eq. (\ref{eq:S1S2Dec}) the term $I1$
can be identified with the structure function of column density
(weighted by $\left\langle v^2\right\rangle$, thus
measurable from observational data), the $I2$ term contains the
analogous in terms of velocity, which could be used directly (assuming
isotropy) to obtain the velocity spectral index. The remaining terms
($I3$ and $I4$) are cross-terms, and cross-correlations, respectively,
and they ``contaminate'' our statistics.  Neglecting them is that we
arrived to our definition of MVCs (LE03), we proposed to simply
subtract the structure function of column density from that of the
centroids.  Later on (EL05, OELS06, \citealt{2007MNRAS.381.1733E}) we
tested the retrieval of the velocity spectral index with various data
cubes, magnetohydrodynamic (MHD) simulations and ensembles of
artificially produced fractional Brownian motion cubes (fBms). Our
results show that centroids are only useful to trace the scaling of
turbulent velocity only when this is subsonic, and that the technique
could be pushed to mildly supersonic turbulence (sonic Mach number
$\lesssim 2.5$) with MVCs. More recently \citep{2007MNRAS.381.1733E},
we have found that the non-Gaussianity of both density and velocity
fields (which is characteristic of highly supersonic turbulence) makes the
approximation in eq.~(\ref{eq:S1S2Dec}) inadequate, the reason being
that the fourth order correlations could not be well approximated by
second order ones in that case. As a guideline for observers, in LE03 we 
proposed a necessary condition for centroids to trace the statistics
of velocity (neglecting $I3$ and $I4$). When
$\left\langle\left[S(\mathbf{X_1})-S(\mathbf{X_2})\right]^2\right\rangle/I1
\ll 1$, then velocity centroids trace the velocity scaling (this is
the case of subsonic turbulence), if the condition is only partially
fulfilled (e.g. the ratio $\lesssim 2$) then MVCs would work, while
the other centroids would not (this was the case of weakly supersonic
turbulence simulations). The condition is strictly speaking a function
of the lag $R$, thus one could have regions where it is true and
regions in where it is not, depending on the slope of the structure
functions involved. In fact one should
compare them at the region in which we measure the spectral
index. But, as
a first approach, one can approximate the condition with the maximal
value of the structure functions, leading to $\langle
\tilde{S}^2\rangle/ \langle \tilde{I}^2\rangle \ll
1$, which can be easily obtained from observations.  A more
restrictive criterion (but that requires additional observational
information) is the ratio of the density dispersion to the mean
density ($\sigma_{\rho}/\rho_0$, OELS06), if this is less than unity
centroids work, if not, then centroids are useless to get the velocity
spectral index.

\subsection{Application to SMC data}
\label{smc}

\begin{figure*}[!t]
  \centering
  \includegraphics[width=\columnwidth]{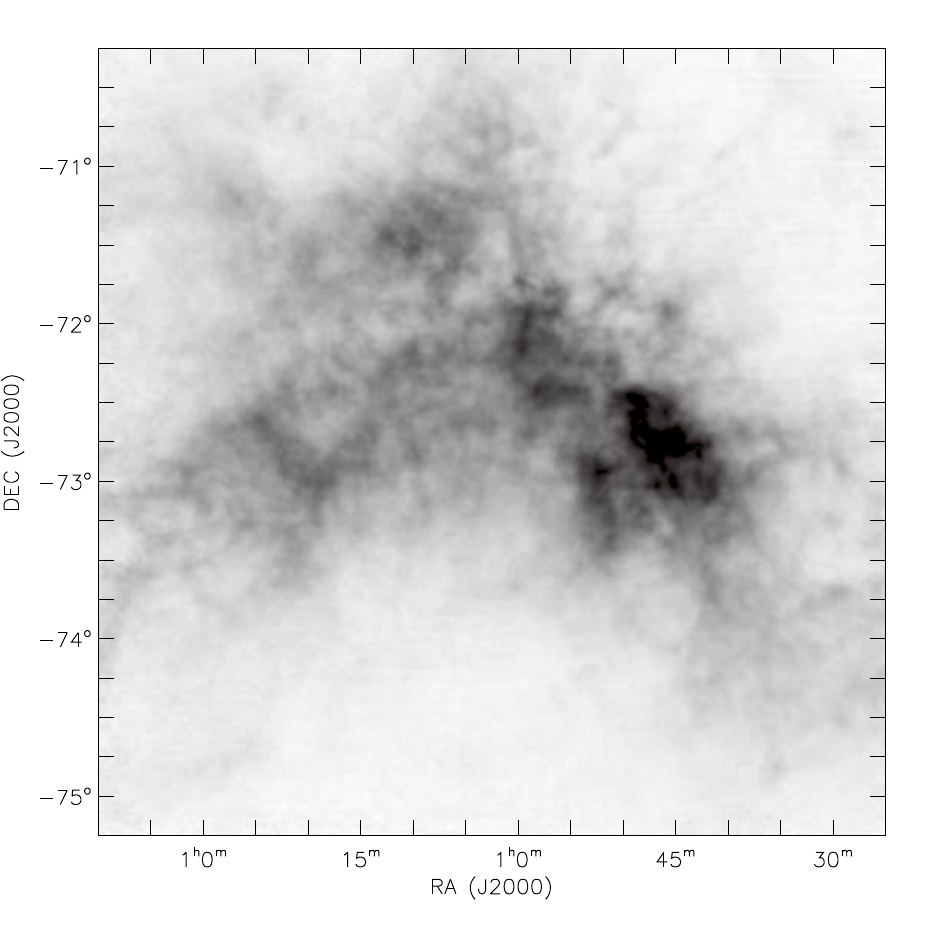}
  \includegraphics[width=\columnwidth]{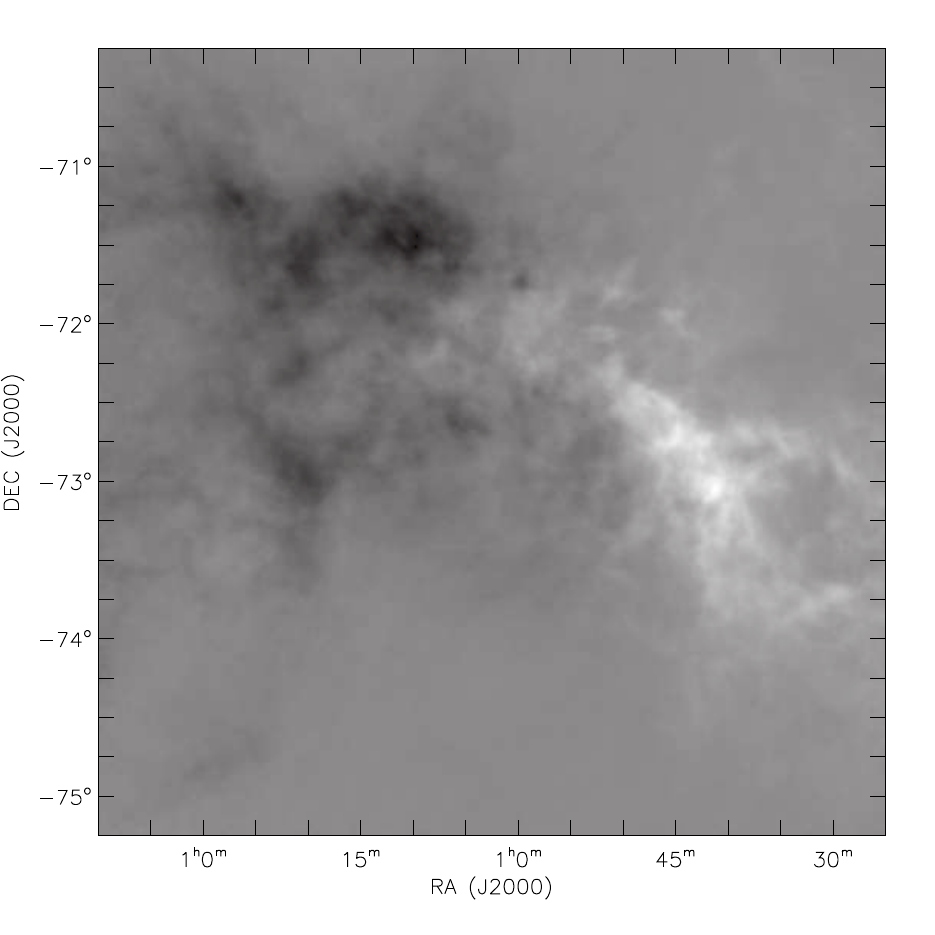}\\
  \includegraphics[width=\columnwidth]{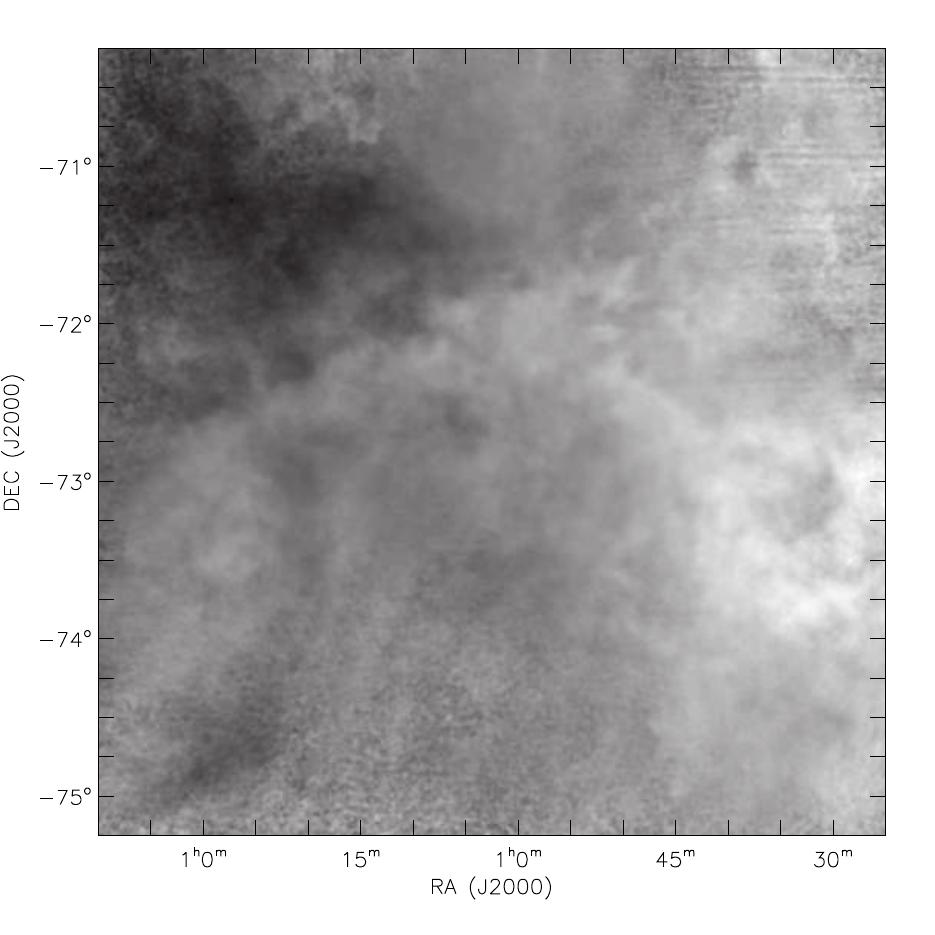}
  \includegraphics[width=\columnwidth]{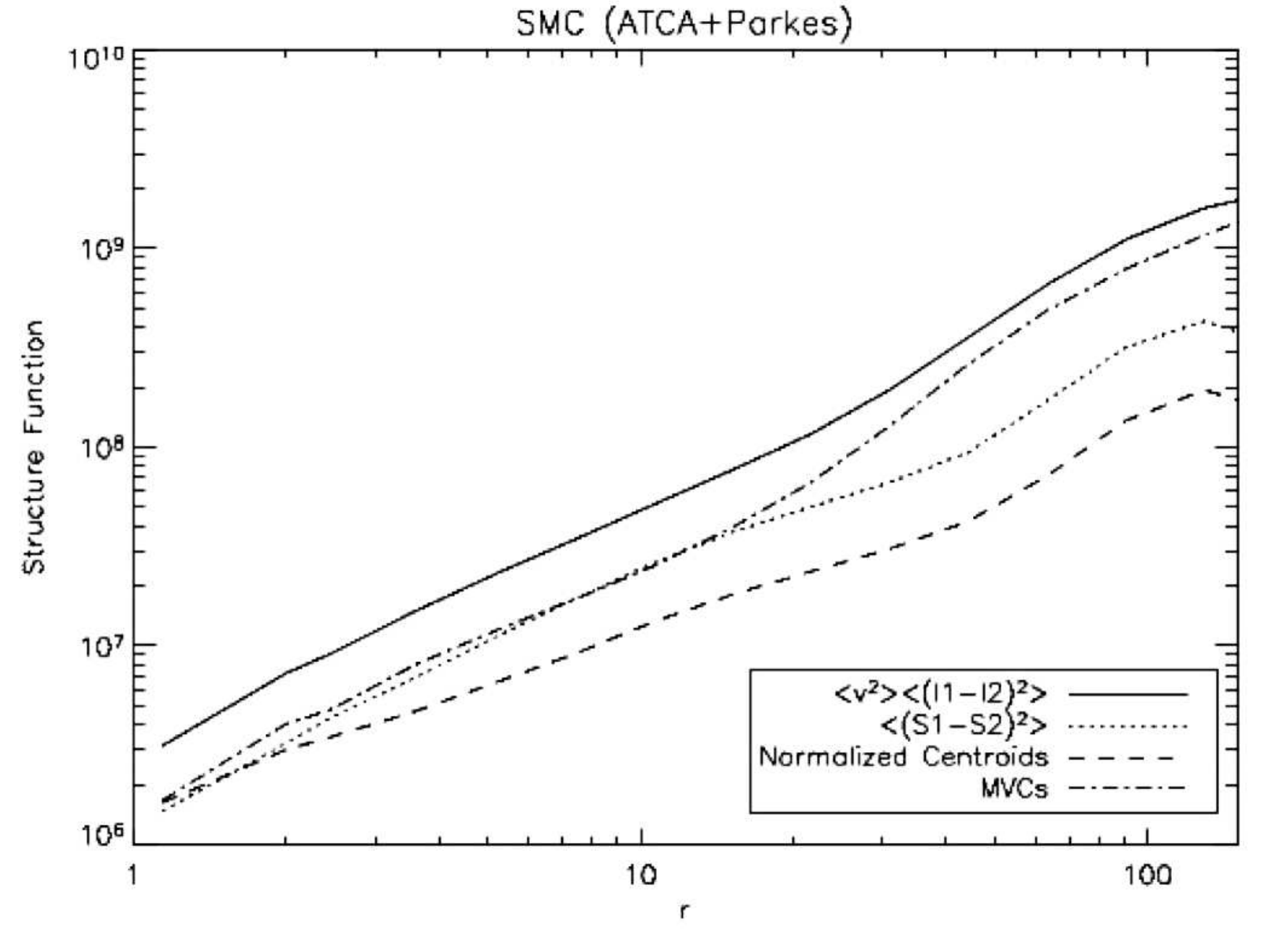}
  \caption{Application to Small Magellanic Cloud observations (data from
    \citealt{1999MNRAS.302..417S}.) 
    {\it Top left}: column density map. {\it Top right} and
    {\it bottom left}: unnormalized velocity
    centroids and normalized velocity centroids, respectively. {\it
      Bottom right}: structure functions of the maps of column density
    and centroids (as indicated in the legend).}
  \label{fig:smc}
\end{figure*}

To illustrate what we describe above, we have taken H{\sc I} 21
cm. data that combines observations with the Parkes telescope and the
Australia Telescope Compact Array (ATCA) interferometer, obtained by
\citet{1999MNRAS.302..417S}.
We obtained from them maps of column density,
velocity centroids (normalized and unnormalized), and MVCs, they are
shown in Figure \ref{fig:smc}.
From both maps of centroids it is evident the global large scale motion
of the SMC, a rotation respect to an axis that forms an angle of $\gtrsim
45\arcdeg$ with respect to the $x$-axis in the maps of the
Figure. On top with this regular motion there is an important turbulent
velocity field.

We have computed the structure function of these quantities, and we show
them in the last panel of Figure \ref{fig:smc}. For this object
$\langle \tilde{S}^2\rangle/ \langle \tilde{I}^2\rangle \sim 0.36$, thus we
should expect centroids {\it not} to trace the scaling of velocity but
rather should be highly affected by density fluctuations. This can be
verified from the figure: $I1=\left\langle
v^2\right\rangle\left\langle (I_1-I_2)^2\right\rangle$
(i.e. $\left\langle v^2 \right\rangle$ times the structure function of
column density) is larger than any other quantity plotted, over the
entire range of scales. The conclusion of this excercise is that to
obtain the spectral index of turbulent velocity in the SMC one should
resort to other techniques, such as VCA
\citep[see][]{2001ApJ...551L..53S}.

\section{The effect of magnetic field: anisotropy in structure and
  correlation functions}
\label{sec:anisotropy}

\begin{figure*}[!t]
  \centering
  \includegraphics[width=\columnwidth]{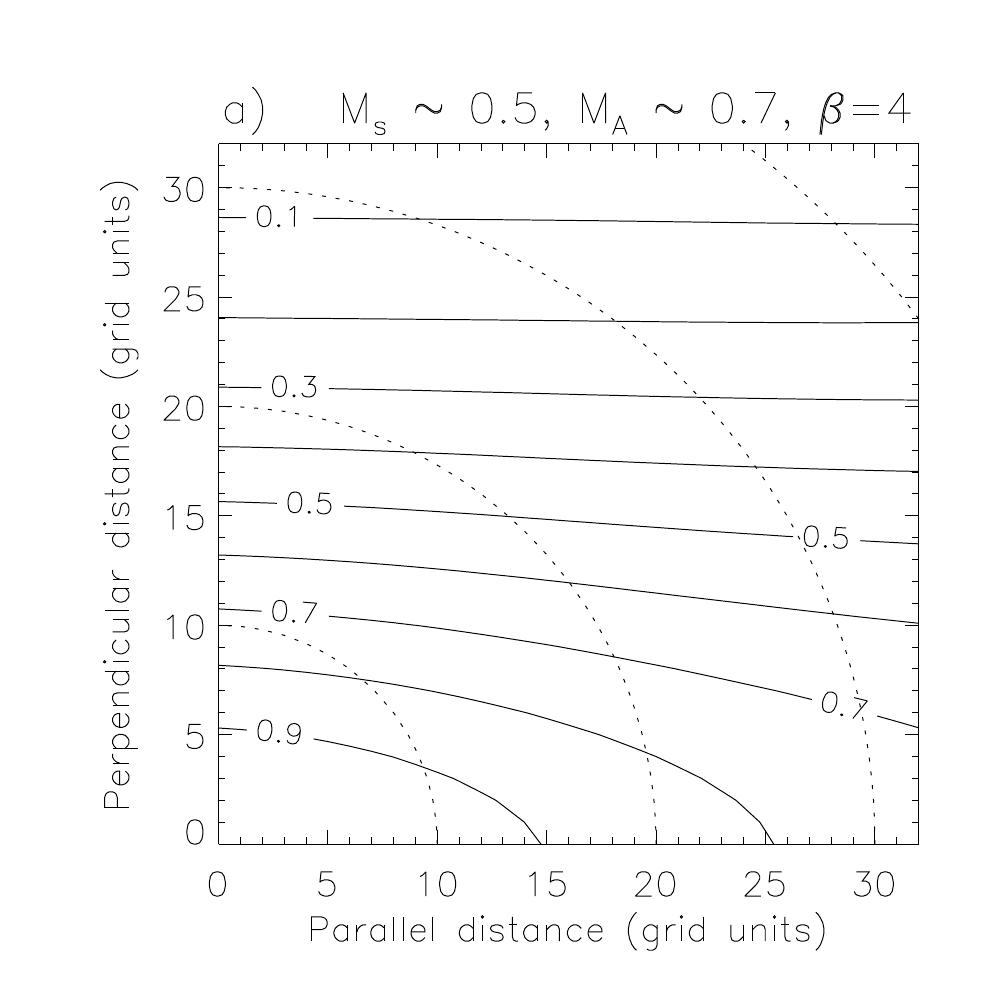}
  \includegraphics[width=\columnwidth]{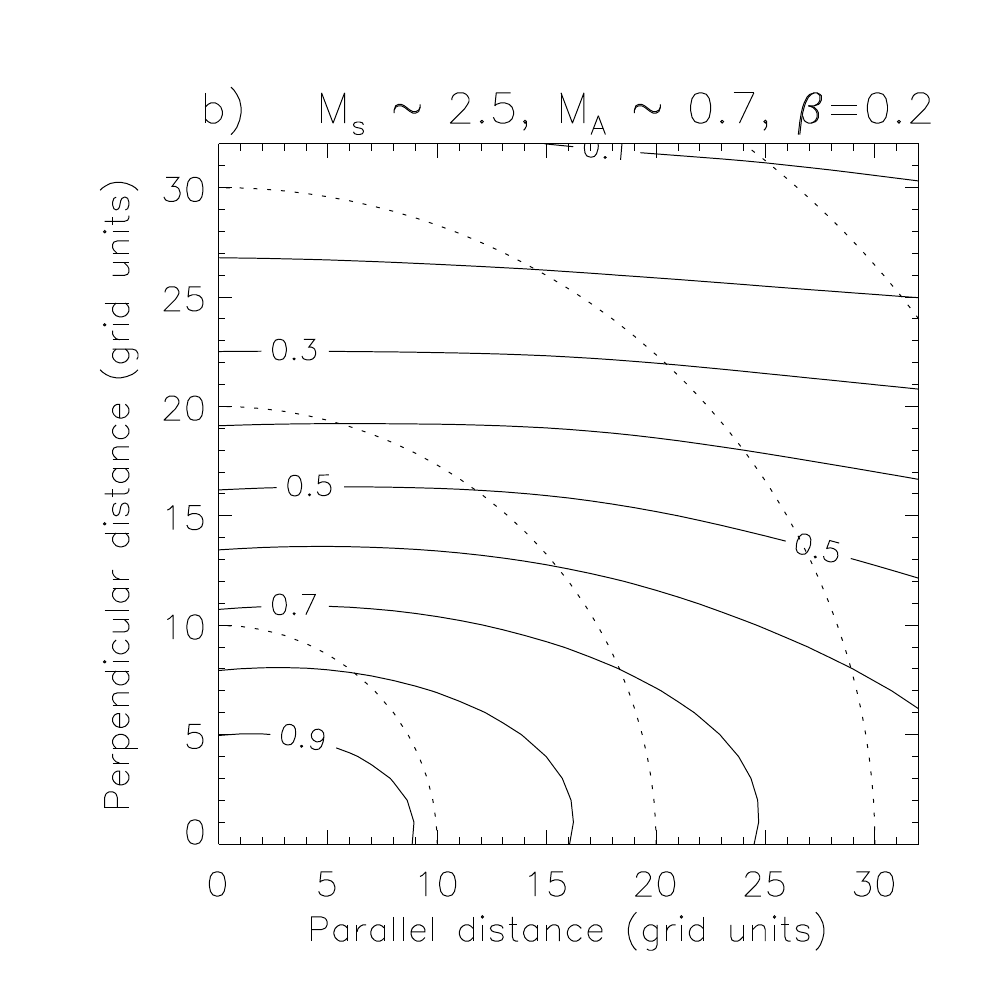}\\
  \includegraphics[width=\columnwidth]{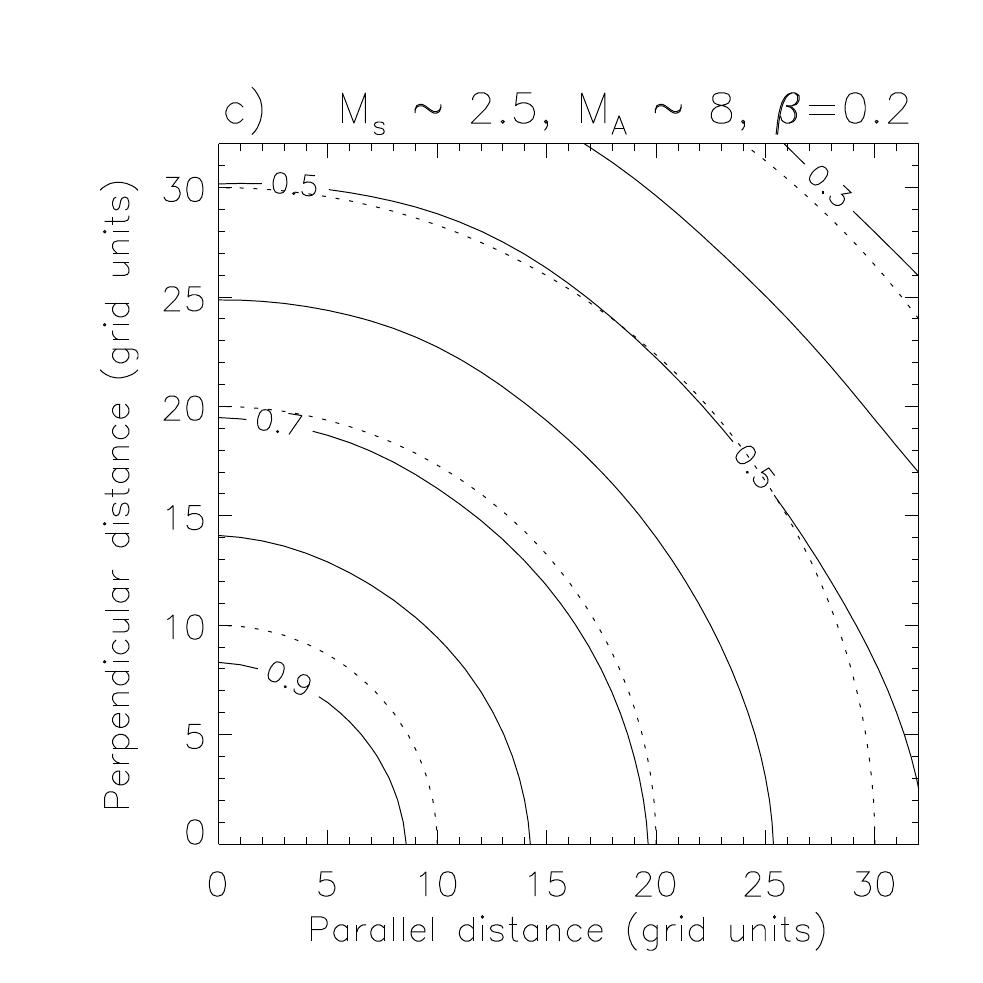}
  \includegraphics[width=\columnwidth]{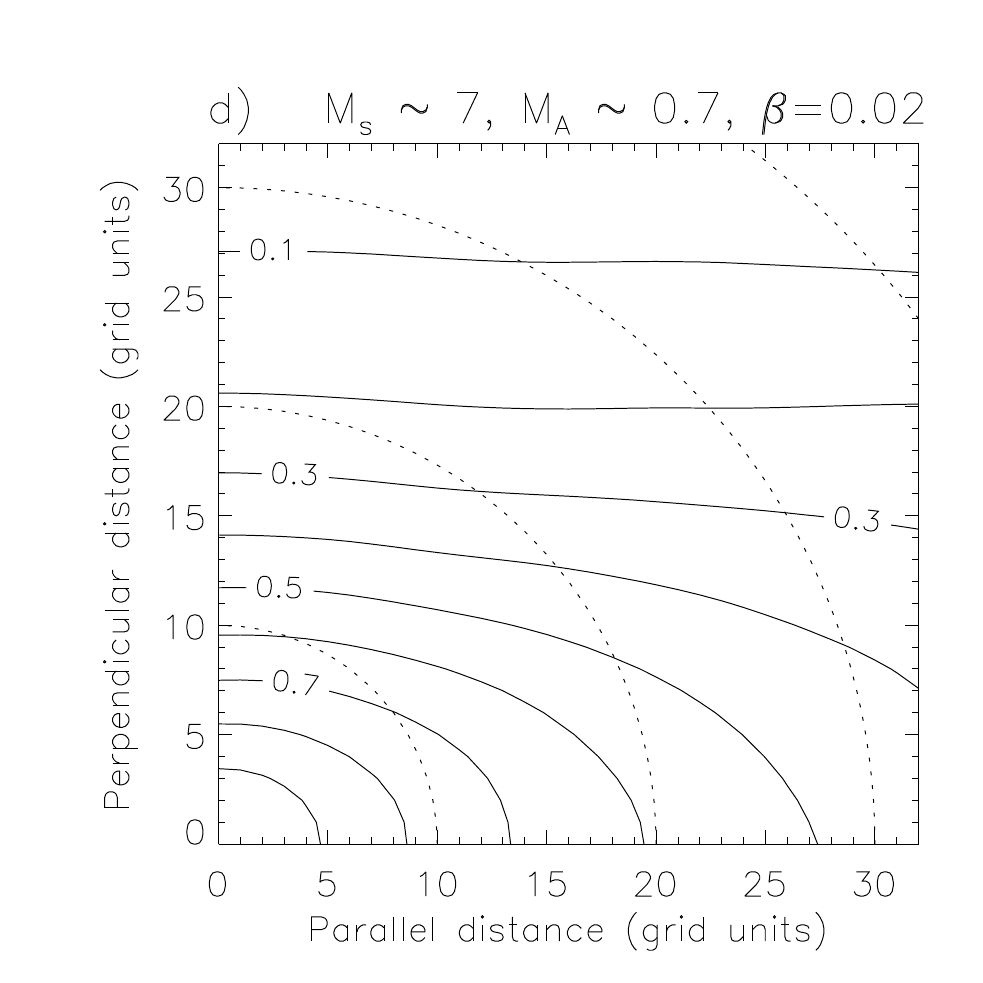}
  \caption{Taken from EL05: Anisotropy in the correlation functions:
    contours of equal correlation for MHD simulations ({\it solid}
    lines). For reference we show isotropic contours as {\it dotted}
    lines. The sonic and Alfv\'{e}n Mach numbers ($\mathcal{M}_s$,
    $\mathcal{M}_A$ respectively), and the plasma $\beta$  are
    indicated in the title of the 
    plots. The anisotropy reveals the direction of the magnetic field
    for all the sub-Alfv\'{e}nic cases, regardless of the sonic Mach
    number. } 
  \label{fig:corr1}
\end{figure*}

\begin{figure*}[!t]
  \centering
  \includegraphics[width=2.\columnwidth]{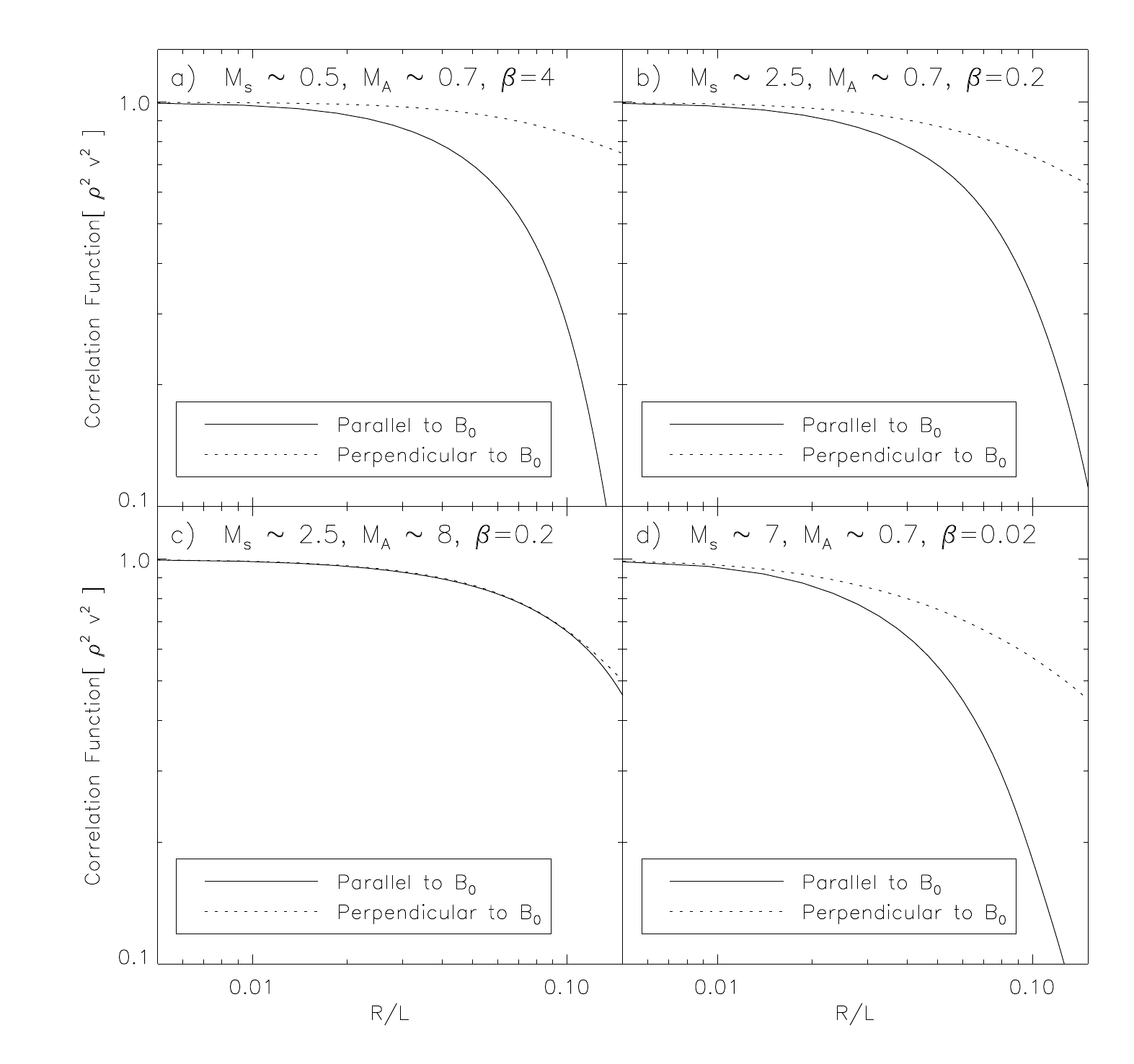}
  \caption{
Taken from EL05: Correlation functions taken in directions parallel
and perpendicular to the mean magnetic field. The anisotropy shows in
the different scale-lengths for the distinct directions. It is
noticeable the little difference in panels ({\it a}), ({\it b}), and
({\it d}). Which correspond to the same ratio $\tilde{B}/B_0$, but
very different sonic Mach numbers. Panel ({\it c}) corresponds to
super-Alfv\'{e}nic ($\tilde{B}>B_0$) turbulence, and the anisotropy is
not evident in the centroid maps. 
}
  \label{fig:corr2}
\end{figure*}

The presence of a magnetic field introduces a preferential direction
of motion that makes the turbulent cascade anisotropic. In a
magnetized plasma, the eddies become elongated in the direction of
their {\it local} magnetic field (see for instance
\citealt*{CLV02}). Fortunately, in spite of being innadequate to recover the
spectral index of velocity, velocity centroids reflect the anisotropic
cascade in sub-Alfv\'enic turbulence (regardless of the sonic Mach
number).
In isotropic turbulence, two-point statistics depend only in the
magnitude of the lag (or wave-number), therefore isocontours of 2D
two-point statistics are circular. In magnetized turbulence, however,
such isocontours become ellipses that are alligned with the
{\it mean} magnetic field. The magnitude of the magnetic field
determines how ellongated they are.
Velocity statistics have been proposed before to study this technique
(e.g. \citealt*{2003MNRAS.342..325E,VOS03}; EL05), but to our
knowledge, they have not been exploited observationally. 

In Figure \ref{fig:corr1} we show (taken from EL05) contours of equal
correlation in  maps of centroids from MHD simulations with different
sonic and Alfv\'en Mach numbers, as well as different plasma $\beta$
(the ratio of gas to magnetic pressures). The parameters of the
simulations are included in the label of each panel, and in more
detail in EL05.

It is evident from the Figure the clear anisotropy present in all the
sub-Alfv\'enic cases, regardless of the large differences in sonic
Mach numbers and/or plasma $\beta$. The only case in which the
anosotropy is not evident is for super-Alfv\'enic turbulence
(panel [{\it c}], with $\mathcal{M_A}\sim 8$). Whether turbulence in the
ISM is typically sub-Alfv\'enic or super-Alfv\'enic is an open debate,
for instance \citet{PJNB04} advocate for a model of supersonic
turbulence in molecular clouds.

A different way to visualize the anisotropy in two-point statistics is
to separate the correlations into perpendicular and parallel to the
{\it mean} (i.e. global) magnetic field. This would mean to plot the
value of the correlations only along the symmetry axes of the ellipses
of equal correlation. We present this in Figure \ref{fig:corr2} (from EL05).

In this case the anisotropy reveals itself as two distinct correlation
lenghts for the parallel and perpendicular direction. The difference
in scale-lengths should in principle reflect a dependence on $(\tilde
B/B_0)^2$, where $\tilde{B}$ is the fluctuating magnetic field and
$B_0$ the mean magnetic field. 
One should have in mind, however, that this method is only sensitive
to the direction of the magnetic field in the plane of the
sky. Turbulence with a strong mean magnetic field oriented along the
line of sight would be indistinguishable from super-Alfv\'enic
turbulence.
It is therefore difficult to determine $(\tilde B/B_0)$
from this type of statistics in supersonic turbulence (where density
fluctations are important). Nonetheless, the results in terms of the
direction of the $B_0$ are robust and could be used when other methods
are not available.

\section{Discussion and Summary}
\label{sec:summary}

We have made a review of our previous work on velocity centroids, and
their application to retrieve the scaling properties of the turbulent
velocity field (i.e. spectral index). 
Additional details can be found
in \citet{2003MNRAS.342..325E}; LE03; EL05; OELS06, and
\citet{2007MNRAS.381.1733E}.
This work on centroids assumes a optically
thin medium with emissivity proportional to density (e.g. H{\sc I}), while
self absorption has been addressed within VCA \citep{LP04} and VCS (Lazarian \& Pogosyan 2006) techniques, which opens a way of formulating the theory of centroids for partially absorbing media. This has not been done yet. 

We also assumed that the emission lines are being used. If the turbulence volume is between the observer and an extended emission source, centroids can be also used with the absorption lines. The present theory assumes that the absorption is in linear regime however. Needless to say, that extend of the emission source of multiple emission sources will determine the spatial coverage of scales that are testable with the centroids. In comparison, the VCS technique can deal with saturated absorption lines and does not depend on the spatial coverage of the data. 

The advantage of centroids compared to the VCA and the VCS techniques is, first of all, the ability of centroids to study magnetic field direction, and, second, ability to study subsonic turbulence more reliably. While the VCA and VCS are also capable of studying subsonic turbulence, the procedures for extracting of information within these techniques are much more complicated. 

The most frequent mistake, we feel, is the use of centroids while dealing with supersonic data (see Miville-Desch{\^e}nes et al. 2003). It is important to understand that when we deal with the multi-phase interstellar medium, e.g. HI, the criterion of being subsonic is the most difficult to be satisfied for the cold medium. Therefore, our research shows that while the application of centroids to HII regions is justified, their applications to find the statistics of velocities in molecular clouds and cold HI cannot deliver reliable spectra. This, as we shown, should not discourage the use of the centroids for studies the direction of magnetic field. The reliability of the latter technique can be tested by comparing the polarization arising from aligned dust (see Lazarian \& Hoang 2007 for a discussion when we can rely on grain alignment to trace magnetic fields) and the results of the statistical anisotropy analysis with centroids. 

The main results of the paper above can be summarized as the following:
\begin{itemize}
\item Centroids maps can be used to trace the spectral index of the
  underyling turbulent velocity in subsonic turbulence. In mildly
  supersonic turbulence (sonic Mach number $\lesssim 2$ can be studied
  with MVCs (LE03, EL05).
\item Two criteria can be used to determine if centroids are useful to
  obtain the velocity spectral index. If
  $\left\langle\left[S(\mathbf{X_1})-S(\mathbf{X_2})\right]^2\right\rangle/I1
  \ll 1$ (see eqs 8, 9) or if $\sigma_{\rho}/\rho_0 < 1$. The first is a necessary
  condition, the latter is more robust measure, but might require
  additional information than what is ussually available from
  spectroscopic observations.

\item We presented an example of the application of velocity
centroids to try retrieving the velocity spectral index from real data
(SMC observations from \citealt{1999MNRAS.302..417S}). The criteria
proposed above was not fulfilled, thus the spectral index should be
studied using an alternate method (such as VCA, see
\citealt{2001ApJ...551L..53S}). We showed how the dominant contribution
to the statistics of centroids are density fluctuations.

\item Two point statistics are anisotropic for sub-Alfv\'enic
  turbulence, both for subsonic and supersonic cases. This anisotropy
  is evident if we plot isocontours of velocity centroids, which
  become elongated and reveal the direction of the {\it mean} magnetic
  field projected onto the plane of the sky. This anisotropy can be
  used where other magnetic field measures are not available.
\end{itemize}

\acknowledgments

\paragraph{Acknowledgments:} AE acknowledges support from the DGAPA
(UNAM) grant IN108207, from the CONACyT grants 46828-F and 61547, and
from the ``Macroproyecto de Tecnolog\'\i as para la Universidad de la
Informaci\'on y la Computaci\'on'' (Secretar\'\i a de Desarrollo
Institucional de la UNAM). AL acknowledges the support from the NSF grant AST 0808118 and the NSF Center for Magnetic Self-Organization in Astrophysical and Laboratory Plasmas.

\end{document}